\documentclass{moriond}

\usepackage[utf8x]{inputenc}
\usepackage{amsmath}
\usepackage{amssymb}
\usepackage{xspace}

\newcommand{\cp}{\ensuremath{\mathcal{CP}}\xspace}
\newcommand{\SM}{{\text{SM}}}

\newcommand{\kgamma}{\ensuremath{\kappa_\gamma}\xspace}
\newcommand{\kg}{\ensuremath{\kappa_g}\xspace}

\newcommand{\cv}{\ensuremath{c_V}\xspace}
\newcommand{\ct}{\ensuremath{c_t}\xspace}
\newcommand{\cttilde}{\ensuremath{\tilde c_{t}}\xspace}

\newcommand{\Photo}{\includegraphics[height=35mm]{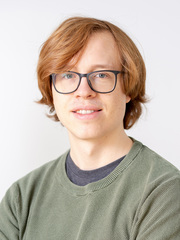}}

\begin{document}

\begin{small}
\hfill DESY 21-065
\end{small}
\vspace*{3.7cm}
\title{Indirect \cp probes of the Higgs--top-quark interaction at the LHC}

\author{ Henning Bahl }

\address{DESY, Notkestraße 85,\\
D-22607 Hamburg, Germany}

\maketitle\abstracts{
The precise determination of the Higgs boson \cp properties is among the most important goals for existing and future colliders. In this work, we evaluate existing constraints on the \cp nature of the Higgs interaction with top quarks taking into account all relevant inclusive and differential Higgs boson measurements. We study the model dependence of these constraints by allowing for deviations from the SM predictions also in the Higgs couplings to massive vector boson, photons, and gluons. Additionally, we evaluate the future prospects for constraining the \cp nature of the top-Yukawa coupling by total rate measurements. In this context, we propose an analysis strategy for measuring $tH$ production at the HL-LHC without relying on assumptions about the Higgs \cp character.
}


\section{Introduction}

The discovery of a new particle which is consistent with the predictions for the Standard Model (SM) Higgs boson marked a milestone for particle physics. The precise determination of its properties is one of the main tasks for future LHC and High-Luminosity LHC (HL-LHC) runs especially in light of the absence of any conclusive hint for beyond the SM (BSM) physics.

One important part of this program --- particularly relevant in the context of searching for new sources of \cp violation in order to explain the baryon asymmetry of the Universe --- is the determination of the Higgs boson's \cp properties. Experimental studies already exclude the possibility of the Higgs boson being a \cp-odd state. The possibility of the Higgs boson being a \cp-admixed state is, however, far less constrained.

The focus of the present study\cite{Bahl:2020wee} is the by magnitude largest Higgs--fermion--fermion interaction: the top-Yukawa interaction. While the \cp properties of the top-Yukawa coupling can also be constrained by electric dipole measurements,\cite{Fuchs:2020uoc} we concentrate on the constraints imposed by existing LHC measurements (i.e.\ by signal strength and differential Higgs boson measurements).

At colliders, \cp-violating couplings can be constrained directly by measuring \cp-odd observable. Measuring a non-zero value for such an observable would directly imply the presence of \cp violation. While proposals for \cp-odd observables targeting the top-Yukawa coupling exists, their measurement is experimentally challenging. \cp violation in the Higgs--top-quark interaction, however, also induces deviations from the SM in \cp-even observables. While a deviation from the SM in a \cp-even observable is not guaranteed to be caused by \cp violation, a systematic investigation of indirect constraints is still a powerful method to narrow down the available parameter space for \cp violation.

In the present work, we perform a fit to derive bounds on a \cp-violating top-Yukawa coupling following the indirect approach. We take into account all available inclusive and differential Higgs boson measurements. Based upon the results of this fit, we point out that a measurement of single top quark associated Higgs production independent of the Higgs \cp character would significantly enhance the sensitivity to a \cp-violating top-Yukawa coupling. We then propose a strategy for performing such a measurements at the HL-LHC focusing on the Higgs boson decay to two photons.


\section{Current constraints}
\label{sec:current}

We parameterize BSM effects in the Higgs-boson interaction with top quarks in the form
\begin{align}\label{eq:topYuk_lagrangian}
\mathcal{L}_\text{yuk} = - \frac{y_t^\SM}{\sqrt{2}} \bar t \left(\ct + i \gamma_5 \cttilde\right) t H,
\end{align}
where $y_t^\SM$ is the SM top-Yukawa coupling, \ct rescales the \cp-even top-Yukawa coupling ($\ct = 1$ in the SM), and \cttilde introduces a \cp-odd top-Yukawa coupling ($\cttilde = 0$ in the SM). Additionally, we introduce the parameter \cv rescaling the Higgs couplings to massive vector bosons. To parameterize the effect of additional BSM particles affecting the $H\to \gamma\gamma$ and $gg\to H$ processes, our fit, moreover, allows to treat the Higgs--gluon--gluon and the Higgs--photon--photon couplings as free parameters --- \kg and \kgamma, respectively.

The most relevant processes to constrain the modified top-Yukawa coupling of Eq.~(\ref{eq:topYuk_lagrangian}) are Higgs production via gluon fusion, the Higgs decay to two photons, $Z$-boson associated Higgs production, and top-associated Higgs production. Note that three different sub-processes contribute to top-associated Higgs production: $t\bar tH$, $tWH$, and $tH$ production. While $tWH$ and $tH$ are negligible in the SM, their cross-section can be significantly enhanced in the presence of a \cp-violating contribution to the top-Yukawa coupling. We do not consider constraints arising from processes involving a virtual Higgs boson (e.g.\ $t\bar t$ or $t\bar t t \bar t$ production).

We perform the global fit of this model to all relevant Higgs boson measurements using \texttt{HiggsSignal}\cite{Bechtle:2013xfa,Bechtle:2014ewa,Bechtle:2020uwn}. \texttt{HiggsSignal} includes the latest Higgs rate measurements from ATLAS and CMS as well as the $p_T$-binned simplified template cross-section measurements for the process $pp\to Z H, H\to b \bar b$.

\begin{figure}\centering
\begin{minipage}{0.45\textwidth}
\includegraphics[width=\textwidth]{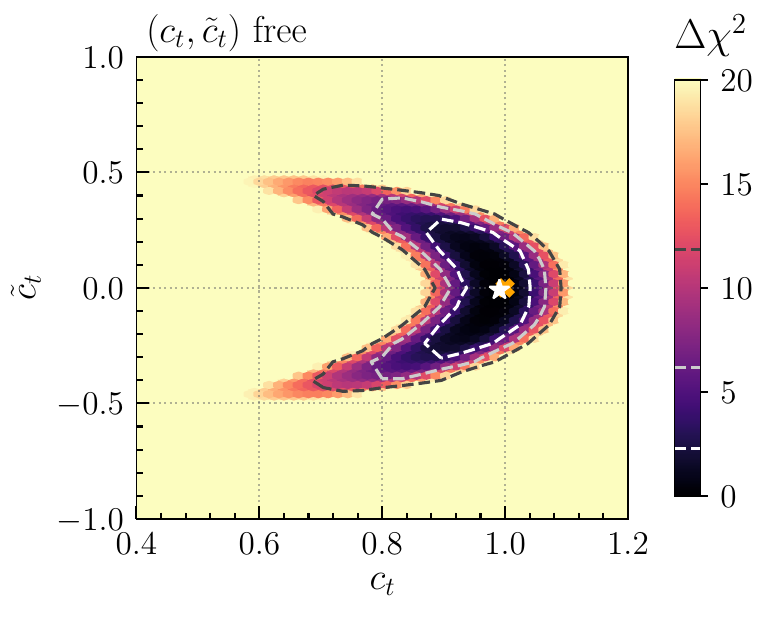}
\end{minipage}
\begin{minipage}{0.45\textwidth}
\includegraphics[width=\textwidth]{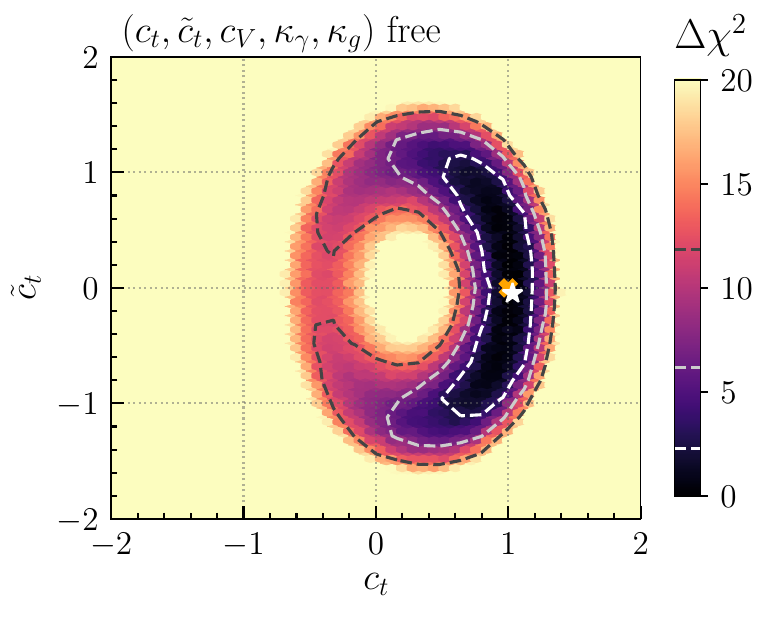}
\end{minipage}
\caption{Fit results in the $(\ct, \cttilde)$ parameter plane. The best fit point is marked by white star; the SM point. by a orange cross. The color encodes the profiled $\Delta\chi^2$ distribution of the fit. The white, light-gray, and dark-gray contour lines represent the $1\,\sigma$, $2\,\sigma$, $3\,\sigma$ confidence regions, respectively. \textit{Left:} Fit results for a simplified two-dimensional model, in which $\cv = 1$ is assumed and \kg and \kgamma are calculated as functions of \ct and \cttilde. \textit{Right:} Fit results for the full five-dimensional model.}
\label{fig:fit_results}
\end{figure}

In Fig.~\ref{fig:fit_results}, we show two exemplary results of the described fit. The left plot shows the $\Delta\chi^2$ distribution in the $(\ct,\cttilde)$ plane for a simplified model in which $\cv = 1$ is assumed and \kg and \kgamma are calculated as functions of \ct and \cttilde. In this model, \cttilde is constrained to the interval $[-0.3,0.3]$ at the $1\,\sigma$ level. The most relevant constraints arise through measurements of Higgs production via gluon fusion and the Higgs decay into two photons. While the rate of Higgs production via gluon fusion would be equal to the SM value for $\ct^2 + 9/4 \cttilde^2 \simeq 1$, the region of $\ct \lesssim 0.7$ is excluded by the constraints arising from the Higgs decay into two photons.

In the right plot of Fig.~\ref{fig:fit_results}, the fit results are shown for the full five-dimensional model in which \ct, \cttilde, \cv, \kg, and \kgamma are treated as free parameters. As a consequence of floating \kg and \kgamma, the $1\,\sigma$ allowed \cttilde range is significantly enlarged to $[-1.1,1.1]$ with respect to the two-dimensional model. The most relevant constraints in this model are rate measurements of top-associated Higgs production --- constraining the fit to an ellipsis where the top-associated Higgs production rate ($t\bar tH + t H + tWH$) is close to its SM prediction. The region of negative \ct is disfavored by $Z$-associated Higgs production measurements as well as measurements able to at least partly disentangle the different contributions to top-associated Higgs production (i.e. disentangling $tH$ from $t\bar t H$ production).


\section{Future constraints}
\label{sec:future}

In the future, the constraints on a \cp-violating Higgs--top-quark interaction can be tightened by either including more kinematic information,\cite{Sirunyan:2020sum,Aad:2020ivc,CMS:2020dkv,Sirunyan:2021fpv} by performing measurements of \cp-odd observables, or by future rate measurements.

\begin{figure}\centering
\begin{minipage}{0.45\textwidth}
\includegraphics[width=\textwidth]{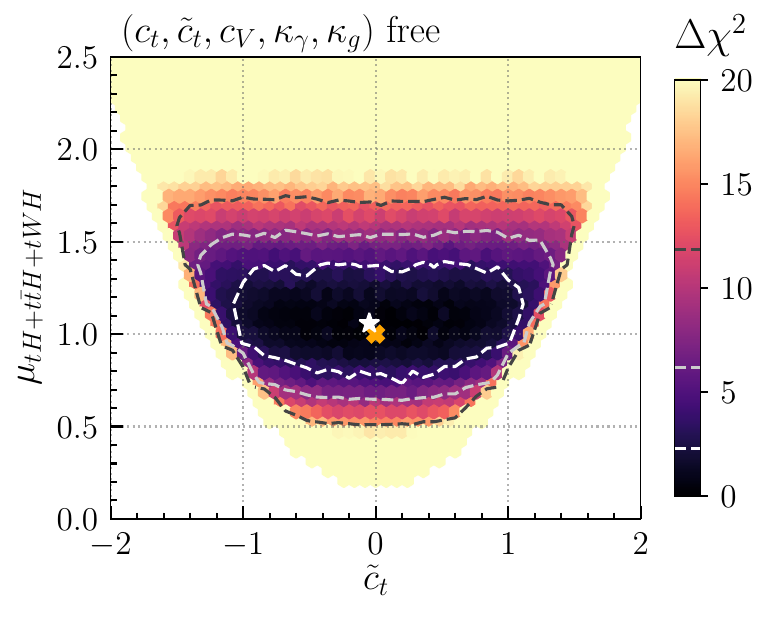}
\end{minipage}
\begin{minipage}{0.45\textwidth}
\includegraphics[width=\textwidth]{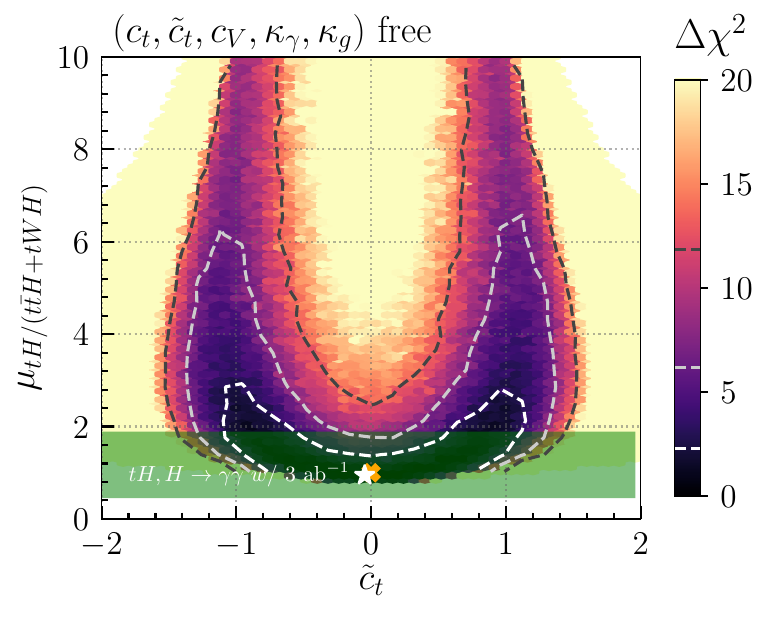}
\end{minipage}
\caption{\textit{Left:} Signal strength for combined top-associated Higgs production for the five-dimensional model in dependence of \cttilde. The color encodes the profiled $\Delta\chi^2$ distribution of the fit. The white, light-gray, and dark-gray contour lines represent the $1\,\sigma$, $2\,\sigma$, $3\,\sigma$ confidence regions, respectively. \textit{Right:} Same as left plot but the signal strength for $tH$ Higgs production divided by the signal strength for $t\bar t H$ and $tWH$ production is shown. The green band indicates the impact of the projected $tH$ measurement with $3\,\text{ab}^{-1}$ at the HL-LHC.}
\label{fig:topH_signal_strengths}
\end{figure}

The most promising target for future rate measurements constraining the Higgs \cp character is top-associated Higgs production. Often a combination of the three channels is measured. In the left plot of Fig.~\ref{fig:topH_signal_strengths}, the cross section for the sum of all three channels (normalized to the SM prediction) is shown as a function of \cttilde as predicted by the fit described in Sec.~\ref{sec:current}. The dependence of $\mu_{tH+t\bar tH+tWH}$ on \cttilde is approximately flat indicating that even a future more precise measurement of $\mu_{tH+t\bar tH+tWH}$ will not allow to tighten the constraints on \cttilde. If it would instead be possible to disentangle $tH$ production from $t\bar tH$ and $tWH$ production, the bounds on \cttilde could be improved. This is evident in the right plot of Fig.~\ref{fig:topH_signal_strengths} displaying the SM-normalized $tH$ over $t\bar t H + tWH$ cross section ratio as a function of \cttilde.

In order to make a $tH$ rate measurement applicable for constraining the \cp nature of the Higgs boson, it should should be ensured that the measurement is independent of the Higgs \cp nature. While e.g.\ the rapidity difference between the leading $b$ jet and the leading non-$b$ jet is a good discriminator between $tH$ and $t\bar tH+tWH$ production, cuts based on this observable introduce a large dependence of the measurement on the \cp character of the top-Yukawa coupling. Using instead the geometric mean of the leading jet and the Higgs boson rapidities, $tH$ and $t\bar tH+tWH$ production can be disentangled equally well without inducing a relevant dependence on the Higgs \cp character. Focusing on the Higgs decay to two photons, this strategy has been used to obtain a projection for a $tH$ measurement at the HL-LHC:\cite{Bahl:2020wee} the $tH$ signal strength is projected to be below $2.21$ at 95\% CL assuming SM-like data and using $3\,\text{ab}^{-1}$ (a limit five times stronger than the current strongest limit\cite{Aad:2020ivc}). The constraint on \cttilde which would be imposed by such a measurement is indicated by the green band in the right plot of Fig.~\ref{fig:topH_signal_strengths}.


\section{Conclusion}

In the present article, we investigated the constraints on a \cp-violating Higgs--top-quark interaction imposed by current LHC data. Using all relevant inclusive and differential Higgs-boson measurements, we performed a global fit in various models.

In the simplest model, the \cp-odd part of the top-Yukawa coupling is strongly constrained by measurements of Higgs production via gluon fusion and the Higgs decay to two photons. Varying additionally the Higgs boson coupling to massive vector bosons as well as to gluons and photons, we found a sizable \cp-odd top-Yukawa coupling to be still allowed with the strongest constraints being imposed by measurements of top-associated and $Z$-associated Higgs production.

As a next step, we discussed the future potential of total rate measurements to tighten the constraints on a \cp-odd top-Yukawa coupling. Since a more precise determination of top-associated Higgs production combining the $tH$, $t\bar tH$, and $tWH$ channels would not result in improved bounds on a \cp-odd top-Yukawa coupling, it will be crucial to disentangle $tH$ from $t\bar tH+tWH$ production without relying on assumptions on the Higgs \cp character. A large deviation of the $tH$ signal strength from the SM would be a strong hint for \cp violation in the Higgs--top-quark interaction.


\section*{Acknowledgments}

I would like to thank Philip Bechtle, Sven Heinemeyer, Judith Katzy, Tobias Klingl, Krisztian Peters, Matthias Saimpert, Tim Stefaniak, and Georg Weiglein for collaboration on the work presented here. H.B. acknowledges support by the Deutsche Forschungsgemeinschaft (DFG, German Research Foundation) under Germany’s Excellence Strategy — EXC 2121 “Quantum Universe” — 390833306.


\end{document}